\documentclass[11pt]{article}
\pdfoutput=1

\usepackage{cite}
\usepackage{amssymb}
\usepackage{graphicx}

\newcommand{\ra}{\rightarrow}
\newcommand{\lsim}{\lesssim}
\newcommand{\gsim}{\gtrsim}

\begin{document}

\vspace*{1cm}

\begin{center} 
{\large\bf Dirac Gauginos in Supersymmetry --
Suppressed Jets $+$ MET Signals: \\ A Snowmass Whitepaper}
\\[1cm] 
Graham D. Kribs$^{1,2}$ and Adam Martin$^{3,4}$ \\[10mm]
$^1$\emph{School of Natural Sciences, Institute for Advanced Study, 
    Princeton, NJ 08540} \\[2mm]
$^2$\emph{Department of Physics, University of Oregon,
    Eugene, OR 97403} \\[2mm]
$^3$\emph{PH-TH Department, CERN, CH-1211 Geneva 23, Switzerland} \\[2mm]
$^4$\emph{Department of Physics, University of Notre Dame, Notre Dame, 
    IN 46556} \\[1cm]

\end{center}

\normalsize

\begin{abstract}

We consider the modifications to squark production in the presence 
of a naturally heavier Dirac gluino.  
First generation squark production 
is highly suppressed, providing an interesting but challenging signal 
find or rule out.  
No dedicated searches for supersymmetry
with a Dirac gluino have been performed, however a reinterpretation of a
``decoupled gluino'' simplified model suggests the bounds on a common
first and second generation squark mass is much smaller than in the MSSM:  
$\lsim 850$~GeV for a massless LSP, and \emph{no bound} for an LSP
heavier than about $300$~GeV\@.  We compare and contrast the squark
production cross sections between a model with a Dirac gluino and one
with a Majorana gluino, updating earlier results in the literature to
a $pp$ collider operating at $\sqrt{s} = 14$ and $33$~TeV\@.
Associated production of squark+gluino is likely very small at 
$\sqrt{s} = 14$~TeV, while is a challenging but important signal
at even higher energy $pp$ colliders.  Several other salient implications 
of Dirac gauginos are mentioned, with some thought-provoking discussion
as it regards the importance of the various experiments planned or proposed.

\end{abstract}

\newpage

\section{Introduction}

Gauginos in weak scale supersymmetry could acquire
dominantly Dirac masses instead of Majorana masses.
Dirac gaugino masses have been considered
long ago \cite{Fayet:1978qc,Polchinski:1982an,Hall:1990hq}
and have inspired more recent model building
\cite{Fox:2002bu,Nelson:2002ca,Chacko:2004mi,Carpenter:2005tz,Antoniadis:2005em,Nomura:2005rj,Antoniadis:2006uj,Kribs:2007ac,Amigo:2008rc,Benakli:2008pg,Blechman:2009if,Carpenter:2010as,Kribs:2010md,Abel:2011dc,Frugiuele:2011mh,Itoyama:2011zi,Unwin:2012fj,Abel:2013kha}
and phenomenology
\cite{Hisano:2006mv,Hsieh:2007wq,Blechman:2008gu,Kribs:2008hq,Choi:2008pi,Plehn:2008ae,Harnik:2008uu,Choi:2008ub,Kribs:2009zy,Belanger:2009wf,Benakli:2009mk,Kumar:2009sf,Chun:2009zx,Benakli:2010gi,Fok:2010vk,DeSimone:2010tf,Choi:2010gc,Chun:2010hz,Choi:2010an,Davies:2011mp,Benakli:2011kz,Kumar:2011np,Davies:2011js,Heikinheimo:2011fk,Fuks:2012im,Kribs:2012gx,GoncalvesNetto:2012nt,Fok:2012fb,Fok:2012me,Frugiuele:2012pe,Frugiuele:2012kp,Benakli:2012cy,Agrawal:2013kha,Hardy:2013ywa,Buckley:2013sca}.
Dirac masses for gauginos of the MSSM requires the model to be extended 
to include a chiral superfields in the adjoint representation of each
gauge group.  Some or all of the gauginos could be Dirac, Majorana, or 
mixed states depending on the model and the mediation of supersymmetry
breaking.  Among the dramatic consequences that have been studied include:
gaugino contributions to scalar masses that are 
``supersoft'' (not log-divergent) \cite{Fox:2002bu};
substantial relief from the supersymmetric flavor problem when the 
low energy model includes an approximate $R$-symmetry \cite{Kribs:2007ac}; 
suppressed EDMs \cite{Hall:1990hq,Kribs:2007ac}; 
heavier Dirac gauginos that are just as naturalness as lighter Majorana
gauginos \cite{Fox:2002bu,Kribs:2012gx,Hardy:2013ywa}; 
the (suppressed) production cross sections of 
colored superpartners at LHC \cite{Kribs:2012gx};
and the absence of some of the historically characteristic signals
of supersymmetry (same sign lepton searches).
There are many other interesting consequences of Dirac (or partially Dirac)
gauginos that we do not have time or space to review, but can be
found in papers cited above.
For this Snowmass white paper, we delineate some ideas for searches
involving Dirac gauginos that provide benchmarks to understand 
the impact of LHC searches thus far, the gaps in the 
searches that persist, and the opportunities for future searches.

One of the important consequences of a Dirac gluino is that 
it can be several times heavier than a Majorana gluino 
with the same degree of naturalness with respect to the
electroweak symmetry breaking scale \cite{Fox:2002bu,Kribs:2012gx}.  
Once a Dirac gluino mass is above roughly $2$-$3$~TeV, 
gluino pair production as well as associated squark-gluino production 
cross sections are negligible at the 8 TeV LHC\@.  
For a Dirac gluino of any mass, several squark production channels vanish 
due to the absence of a ``chirality flipping'' Majorana mass in
$t$-channel exchange, namely $pp \ra \tilde{q}_L\tilde{q}_L$, 
$pp \ra \tilde{q}_R\tilde{q}_R$, etc. 
Other squark production channels that involve the gluino in $t$-channel
exchange, such as $pp \ra \tilde{q}_L\tilde{q}_R$, are suppressed by
$|p|/M_{\tilde{g}}^2$ in the amplitude, where $|p|$ is the momentum
in the propagator.  
This suggests the dominant production mode of colored superpartners is 
$pp \ra \tilde{q}\tilde{q}^*$ for
first (and second, third) generation squarks is through $s$-channel 
gluon exchange.  
The total colored superpartner production cross section is therefore reduced 
by roughly two orders 
of magnitude compared with what is typical in the MSSM -- 
a Majorana gluino roughly equal in mass to the squarks.

\section{Existing Searches}

There are no dedicated searches for supersymmetric models with
a gluino that acquires a Dirac (or ``mixed'' -- Dirac and Majorana) mass. 
Since a Dirac gluino can be a factor of several times heavier 
than a Majorana gluino without additional 
fine-tuning, one interesting scenario to consider is when
the squark masses are generated dominantly from the finite 
contributions from the Dirac gluino.  In this case, 
$M_{\tilde{g}}/M_{\tilde{q}} \simeq 5 \ra 10$. 
For squark masses larger than $500$~GeV, the Dirac gluino is 
sufficiently heavy $\gsim 2.5$~TeV, and with sufficiently
suppressed effects at the 8 TeV LHC, that it is effectively
decoupled from the spectrum.  Note that this is \emph{not true}
of a Majorana gluino of the same mass, due to the structure
of the interactions, in particular, the lower dimension operator
for squark production with a Majorana mass insertion
(that leads to $1/M_{\tilde{g}}$-suppressed effective interactions). 
We can therefore map a modestly heavy Dirac gluino into
the existing ``decoupled gluino'' simplified model searches 
performed by the ATLAS and CMS collaborations. 

Squark production with a decoupled gluino is  
a specific simplified model in which the ATLAS and CMS
collaborations have placed bounds on the cross sections
within the $(M_{\tilde{q}},M_{\rm LSP})$ space.  
Here we mention only the results from the latest analyses, 
namely a jets plus missing energy search strategy at ATLAS using 
$\simeq 20$~fb$^{-1}$ of 8 TeV data at ATLAS \cite{ATLAS_jetsmissing2013}
and CMS \cite{CMS_jetsmissing2013}.  The simplified model
used for each search assumed
the first and second generation
squarks have a common mass $M_{\tilde{q}}$, there is a single
LSP with mass $M_{\mathrm{LSP}}$, and the squarks are assumed 
to decay $100\%$ of the time via $\tilde{q} \ra q + \mathrm{LSP}$.
For a nearly massless LSP, the current ATLAS results rule out 
$M_{\tilde{q}} \lsim 850$~GeV while the CMS results rule 
out $M_{\tilde{q}} \lsim 780$~GeV\@.  
If the LSP mass were $300$~GeV, the ATLAS and CMS analyses 
\cite{ATLAS_jetsmissing2013,CMS_jetsmissing2013} find \emph{no bound}
on the squark mass. 
These results can be contrasted with, for example, the current
bound on the MSSM simplified where a Majorana gluino
is also present in the spectrum.  Consider the case where 
$M_{\tilde{q}} = M_{\tilde{g}}$:  the latest ATLAS result
constrains $M_{\tilde{q}} (= M_{\tilde{g}}) \gsim 1.7$-$1.8$~TeV
for an LSP up to $700$~GeV in mass \cite{ATLAS_jetsmissing2013}.  
Regarding the search for first and second generation squarks with a 
decoupled gluino, the ATLAS note stated \cite{ATLAS_jetsmissing2013}:
``the expected limits for [the decoupled gluino case] do not extend 
substantially beyond those obtained from the previous published ATLAS 
analysis because the events closely resemble the predominant 
$W$/$Z + 2$-jet background, leading the background uncertainties to 
be dominated by systematics.''  
This implies the need for innovative search strategies to uncover 
squark production with suppressed production cross sections.

\section{Motivation for Future Collider Studies}

Clearly, if the first and second generation squarks are 
much lighter than the standard MSSM analyses suggest, 
it is worthwhile to consider what future collider studies
can say about this scenario.  Following our earlier analysis 
from 2012 \cite{Kribs:2012gx}, we consider
the following scenarios:  ``Dirac5'', ``MSSM5'', ``MSSMequal''
detailed in Table~\ref{scenarios-table}.
These are \emph{not} meant to represent to full spectrum
of possibilities or phenomenology associated with
suppressed colored sparticle production.  Instead, 
we are interested to provide a few benchmark examples
of the differences between gluinos having a Dirac mass 
versus a Majorana mass as it pertains to searches for 
(highly) suppressed colored sparticle production. 
\begin{table}
\begin{center}
\begin{tabular}{c|ccc}
                    & Dirac5      & MSSM5       & MSSMequal  \\ \hline
$M_{\tilde{g}}$     & $5$~TeV     & $5$~TeV     & $=M_{\tilde{q}}$ \\
$M_{\tilde{q}}$     & varies      & varies      & $=M_{\tilde{g}}$ \\
$\tilde{q}$         & 1st,2nd gen & 1st,2nd gen & 1st,2nd gen \\
$BR(\tilde{q} \ra q + \mathrm{LSP})$
                    & $100\%$     & $100\%$     & $100\%$  \\
LSP mass            & $0$         & $0$         & $0$  
\end{tabular}
\end{center}
\caption{Simplified models considered in this writeup.  
All masses are in TeV\@.  Sparticles not listed are decoupled.}
\label{scenarios-table}
\end{table}

\begin{figure}
\includegraphics[width=0.5\textwidth]{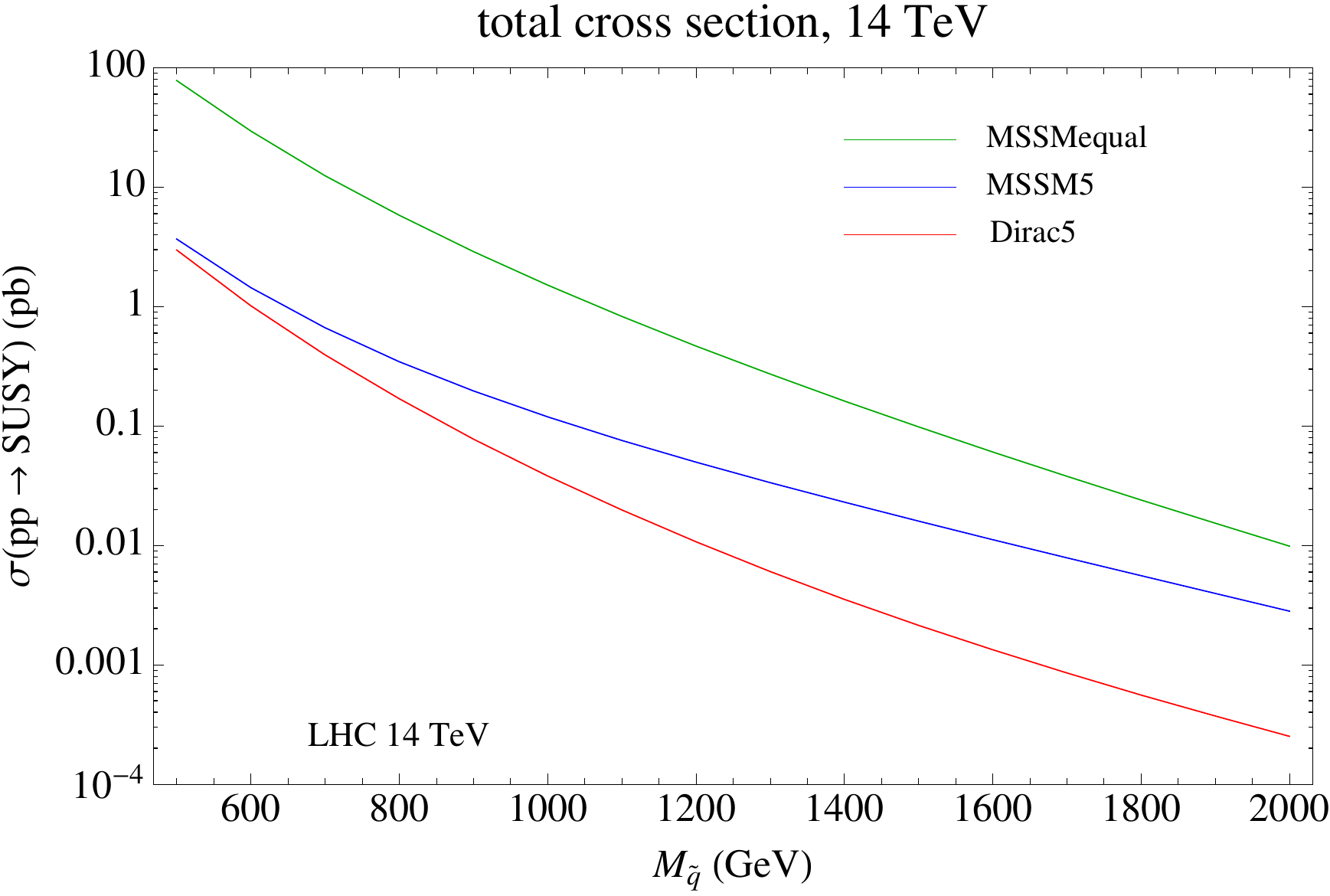} \hfill
\includegraphics[width=0.5\textwidth]{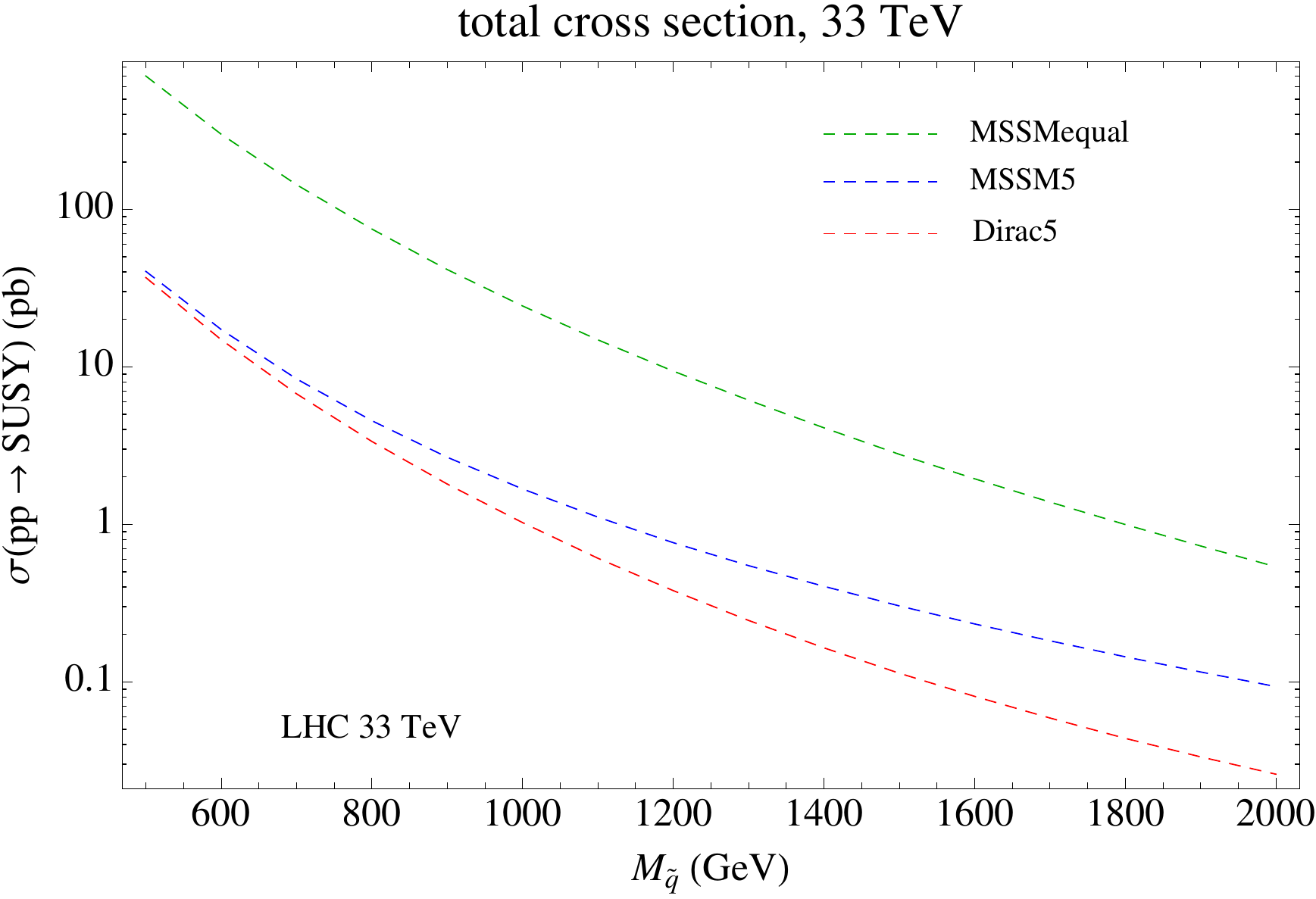} \\
\includegraphics[width=0.5\textwidth]{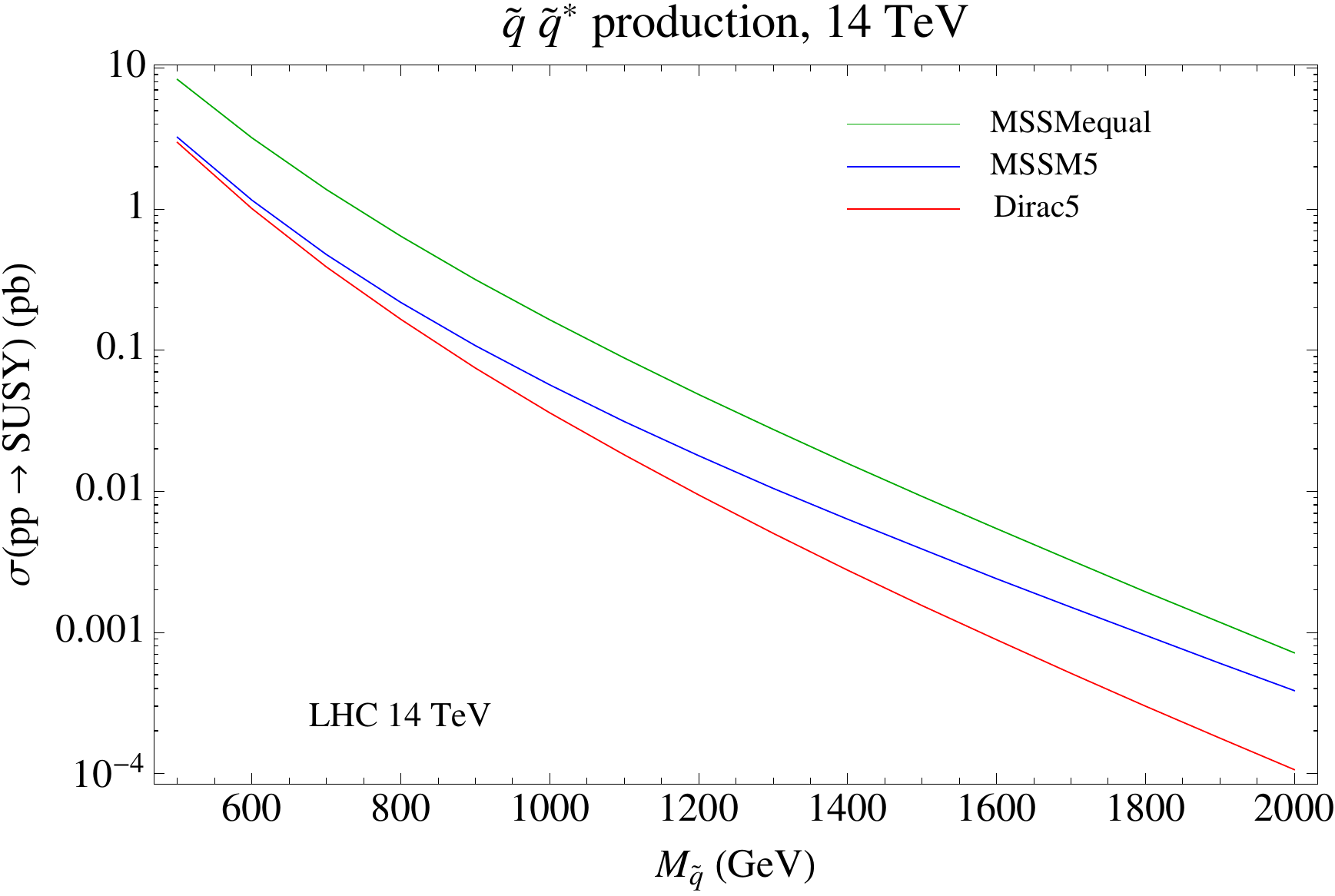} \hfill
\includegraphics[width=0.5\textwidth]{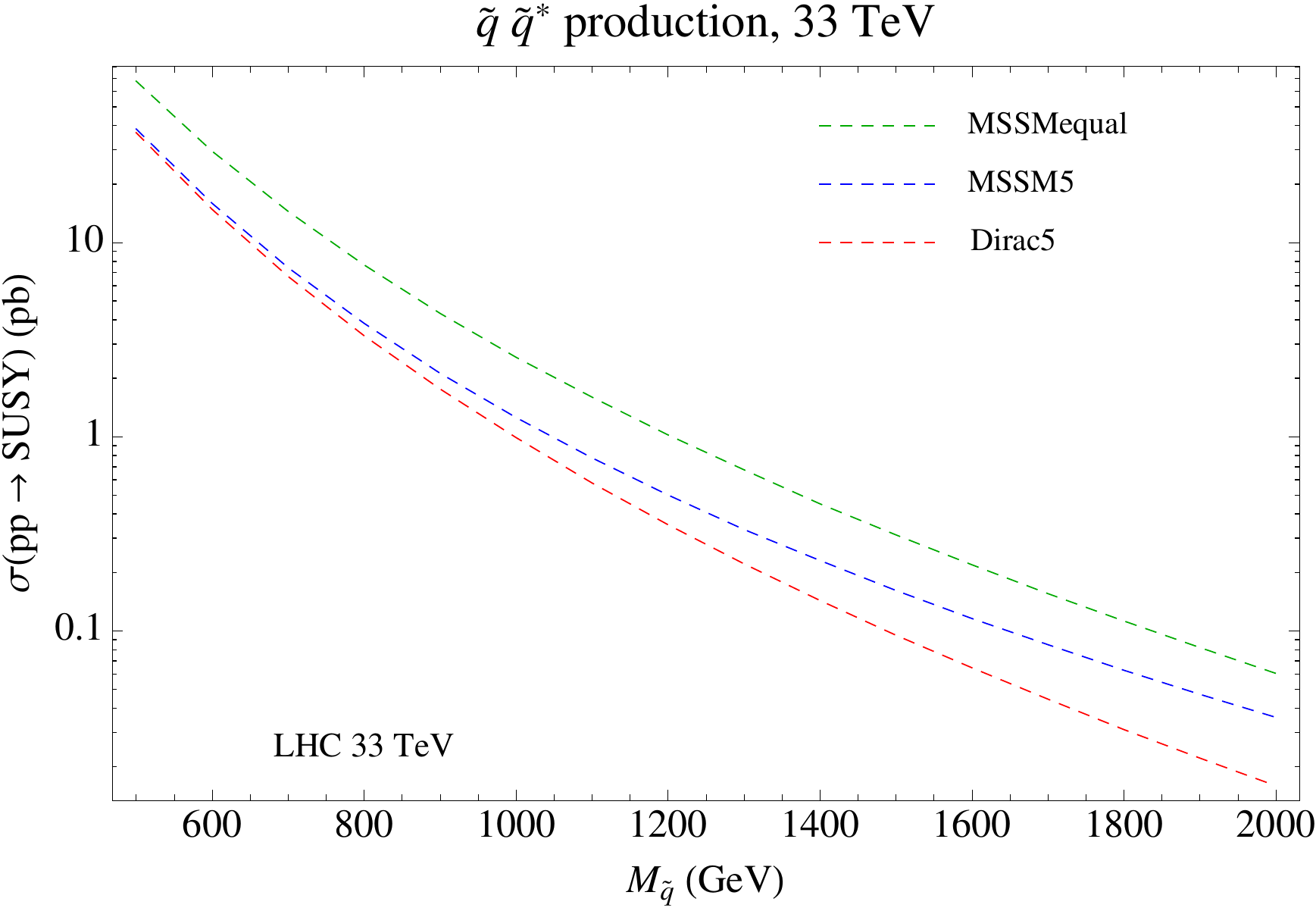} \\
\includegraphics[width=0.5\textwidth]{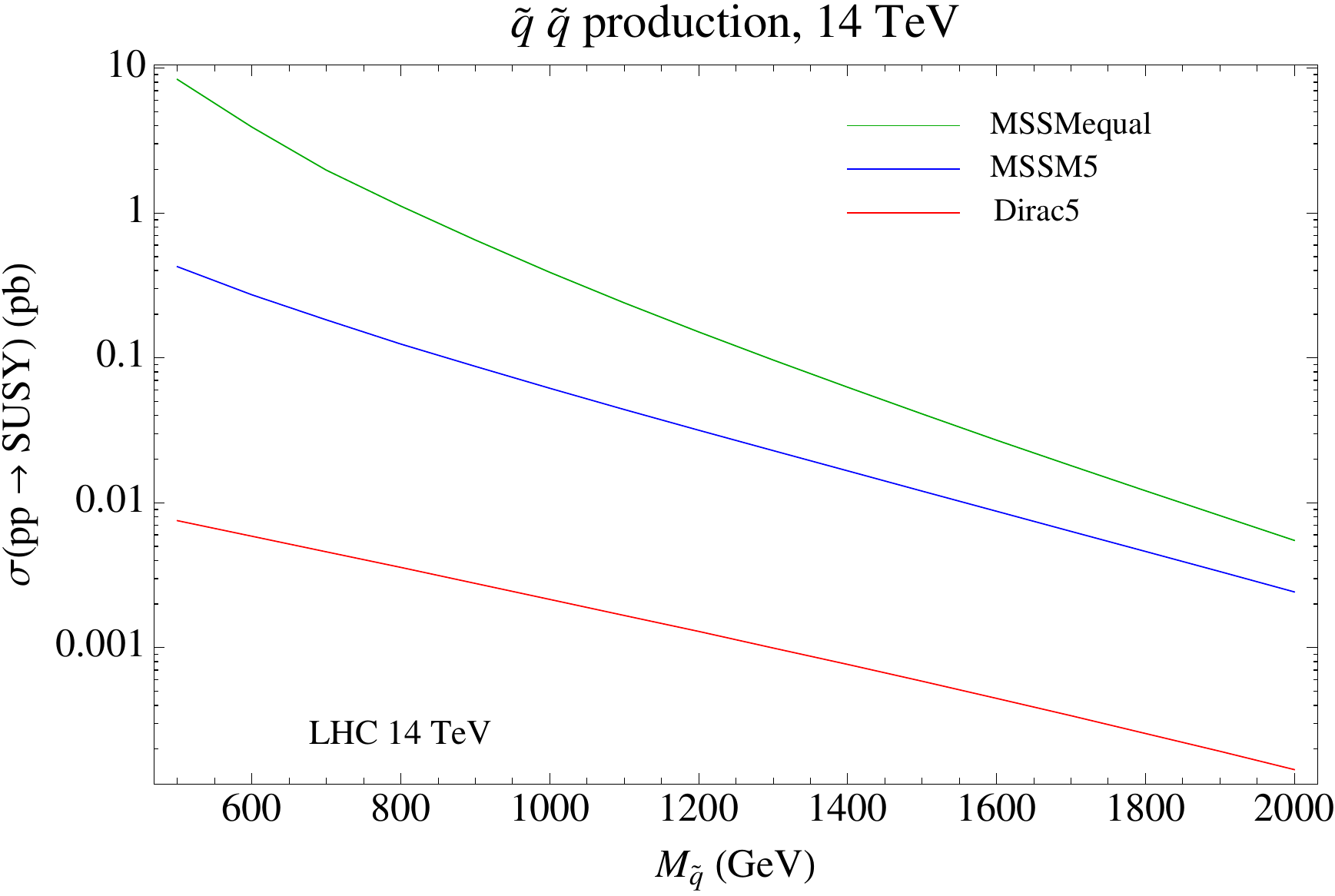}  \hfill
\includegraphics[width=0.5\textwidth]{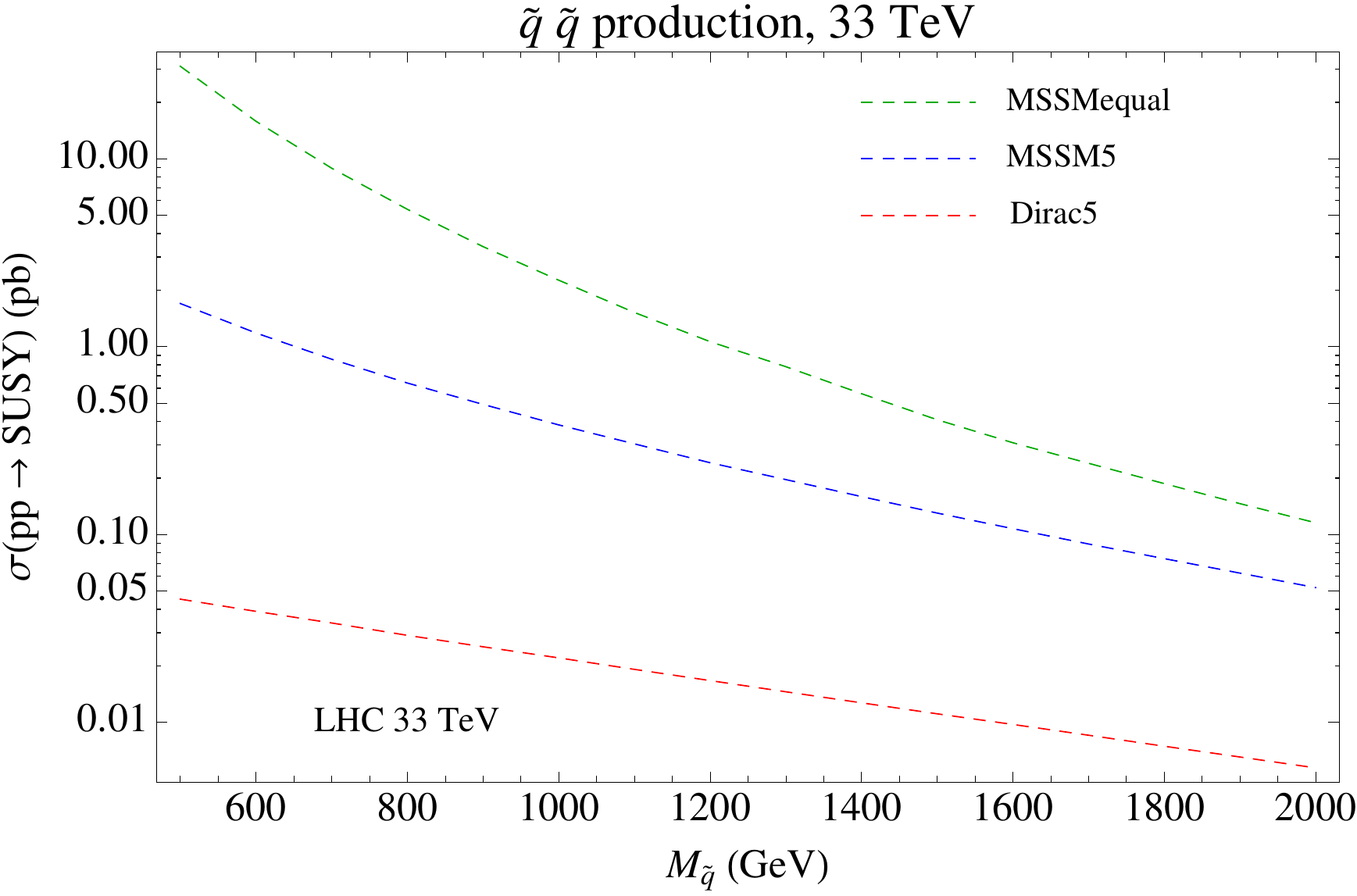}
\caption{Cross sections for the simplified models considered in this writeup.
For squark production, all allowed combinations 
of the first two generations of squarks are summed together.
For total colored sparticle production, both gluino pair production
and gluino-squark associated production are included.
In all results we used MadGraph4 \cite{Alwall:2007st}
at leading order, for LHC operating at $\sqrt{s} = 14$ and $33$~TeV\@.}
\label{fig:crosssections1}
\end{figure}

In Fig.~\ref{fig:crosssections1}, we show the production cross sections
for three quantities:  the total colored sparticle production
(squark and gluino production), the cross section of $\tilde{q}\tilde{q}^*$, 
and the cross section of $\tilde{q}\tilde{q}$.   All allowed combinations 
of the first two generations of squarks are summed together.
In all results we used MadGraph4 \cite{Alwall:2007st}
at leading order, for LHC operating at $\sqrt{s} = 14$ and $33$~TeV\@\footnote{We use CTEQ6L1 parton distribution functions and default factorization and renormalization scales for all simulations}.
The figures clearly show the suppression in cross sections persist
at LHC collider energies of $14$ and $33$~TeV\@.  Notice that
$\tilde{q}\tilde{q}$ production is always subdominant to 
$\tilde{q}\tilde{q}^*$ production for a scenario with a $5$~TeV
Dirac gluino, throughout the squark mass range shown, $M_{\tilde{q}} < 2$~TeV\@.
By contrast, $\tilde{q}\tilde{q}$ production is comparable or dominates
the production cross section of squarks for either scenario involving
a Majorana gluino.

\subsection{Storyboard:  Discovery of suppressed $M_{\tilde{q}} = 1$~TeV
                        at LHC with $\sqrt{s} =14$~TeV and $\simeq 100$~fb$^{-1}$.}

In the spirit of the Irvine ``storyboards'' \cite{lutyslides} for discovery of 
new physics in the next run of the LHC, we consider the possibility
that the LHC has discovered a jets plus missing energy signal consistent 
with first and second generation squark production with squark mass 
$M_{\tilde{q}} = 1$~TeV, but with a highly suppressed cross section 
relative to the expectations from MSSM\@.

There are several investigations one would like to apply to the signal.
The first obvious one is to try to pin down the mass scale of the
squarks and obtain an upper bound on the LSP mass.  
This requires careful examination of the signal kinematic distributions,
e.g.~\cite{Barr:2011xt}.
Searching for accompanying signals, namely in the $n \ge 3$-jet 
categories could uncover evidence for, or absence of, 
an accompanying gluino production signal.
Even if there is no accompanying signals consistent with
a kinematically accessible gluino, we saw above that the 
squark production rates are nevertheless sensitive to a 
kinematically inaccessible Majorana gluino.  This can be seen by
contrasting the squark production rates for the MSSM5 scenario 
against the Dirac5 scenario.  If the experimental data on the
the squark production rate appears to be consistent with just
$\tilde{q}\tilde{q}^*$ production, the signal can be probed for 
consistency with this hypothesis.  For example, by measuring 
the angular distributions of the final state decay products should allow the
experiments to verify the signal is consistent with $s$-channel 
gluon production of $\tilde{q}\tilde{q}^*$ (versus a $t$-channel 
gluino-mediated production of $\tilde{q}\tilde{q}$ with a rate
that happened to match the ``observed'' squark--anti-squark rate). 

If the rate is slightly larger than what is expected from just
$\tilde{q}\tilde{q}^*$ production, there are several possible
culprits.  One is that the gluino is not completely decoupled,
and its effects on $t$-channel exchange are being (slightly) felt.
Another is that the first and second generation squarks are not
precisely degenerate in mass, but these kinematic differences
are not readily observable.

The central goal in this scenario would be to discover the heavy gluino state.
This is where a Dirac gluino becomes much more advantageous
compared with a Majorana gluino (holding the squark production
cross sections roughly the same).  Because the Dirac gluino can be
much lighter without affecting squark production channels, this suggests
associated $\tilde{g}+\tilde{q}$ production can be probed by 33 TeV LHC\@. 
The leading order rates for $\tilde{g}+\tilde{q}$ production in Dirac5 scenario, 
with $M_{\tilde{q}} = 1$~TeV and $M_{\tilde{g}} = 5$~TeV, are 
\begin{eqnarray*}
\sigma(\tilde{q}+\tilde{g}) &\sim& 0.015 \; {\rm fb} \qquad 
  \sqrt{s} = 14 \; {\rm TeV} \\
\sigma(\tilde{q}+\tilde{g}) &\sim& 12 \; {\rm fb} \;\;\;\;\qquad 
  \sqrt{s} = 33 \; {\rm TeV} \, .
\end{eqnarray*}
Clearly, the cross section at the LHC operating at $14$~TeV is 
much too small to be seen at any conceivable integrated luminosity.
However, at the higher center-of-mass energy of $33$~TeV, 
it is at least conceivable to obtain evidence for a heavy gluino
given that the cross section is nearly $3$ orders of magnitude larger.
At even higher energy machines ($100$~TeV), there should be no
difficulty measuring and studying both associated and gluino pair production.

\section{Discussion}

We have focused on the narrow issue of first and second generation
squark production in the presence of a heavy Dirac gluino.
The highly suppressed cross section of lighter squarks 
is an incredibly important signal to find or rule out,
to determine if supersymmetry is realized in this interesting
but non-standard way.  Our quick-and-dirty preliminary investigation
suggests that high energy is more important that high luminosity,
since kinematics may well limit the ability to probe the Dirac
gluino directly.  
This are many more interesting issues that could be explored,
and thus far, have few if any experimental analyses completed:

\begin{itemize}

\item What happens when there is a heavy Dirac gluino and lighter
Majorana electroweak gauginos?  This is an interesting scenario
where the electroweak $D$-term is not suppressed (giving the ordinary
tree-level Higgs mass contribution familiar from the MSSM), 
and does not appear to significantly affect squark 
production \cite{Kribs:2013eua}. 
How much of the same-sign dilepton signal remains when the 
gluino is Dirac while the electroweakinos are Majorana?

\item What happens when the first and second generation,
or even up-type and down-type squarks are not degenerate in mass?
A few interesting recent examples that contains a light second
generation can be found in Ref.~\cite{Mahbubani:2012qq,Galon:2013jba}. 

\item What happens if there is substantial squark mixing, 
e.g., following the $R$-symmetric supersymmetric model 
\cite{Kribs:2007ac,Kribs:2009zy}.
Squark decays to heavy flavor (tops, bottoms, and $\tau$s) become generic.
Are the bounds better in the case of third generation squarks
(given multitude of dedicated analyses for this ``natural'' region)?
How effective (or ineffective) are the standard jets plus missing energy 
strategy?  

\item How is $m_h = 125$~GeV realized in this scenario?
In scenarios with Dirac electroweak gauginos, there are methods 
to raise the tree-level contribution mass (for example, \cite{Fok:2012fb}),
that typically also require moderately heavy stops
($M_{\tilde{t}} \gsim 2$-$3$~TeV).  How easily can the stops be
probed with colliders?  This is a common but important issue 
in this and most other supersymmetric scenarios.

\item ILC implications?  If Dirac electroweak gauginos are
present, the sleptons and Higgs scalars acquire weak-coupling
suppressed contributions to their masses, and thus are generically
light.  In addition, in models with an approximate 
$R$-symmetry \cite{Kribs:2007ac}, there are more ``Higgs-like''
states to uncover, since each Higgs supermultiplet ($H_u,H_d$) 
is paired up with an $R$-supermultiplet partner ($R_u,R_d$)
giving many more scalars and fermions with electroweak 
interactions with the Standard Model.

\end{itemize}

Finally, we should emphasize that in complete models involving
Dirac gauginos, there are a host of indirect methods to probe
the model.  These rely on intensity frontier experiments, 
adding to the motivation that a diverse array of experiments
in particle physics \emph{that can probe new mass scales}
are an essential complement to the energy frontier experiments.

For example, if the low energy theory contains large squark
or slepton mixing, there is the possibility to observe a 
flavor-changing neutral current process at a level that is
expected to be probed by future experiments.  One example is
charged lepton flavor violation.  It was shown in \cite{Fok:2010vk}
that the-then existing bounds from the MEG experiment on $\mu \ra e\gamma$
were only beginning to probe \emph{maximal} lepton flavor violation
in an $R$-symmetric model with Dirac gauginos.  One of the
interesting results is that $\mu \ra e$ conversion experiments
are generically more sensitive to CLFV in $R$-symmetric 
supersymmetry, and happily we expect the Mu2e experiment
and future Project X experiments to probe between four to six
orders of magnitude lower in rate than the best bound today.

\end{document}